# Crystal nucleation and growth in high-entropy alloys revealed by atomic electron tomography

Yakun Yuan[1,2], Saman Moniri[1], Yao Yang[1], Jihan Zhou[1], Andrew Yuan[1], Dennis S. Kim[1], Yongsoo Yang[1], Chenyang Li[3], Wei Chen[3], Peter Ercius[4], Jianwei Miao[1*]

*[1]Department of Physics and Astronomy and California NanoSystems Institute, University of California, Los Angeles, CA 90095, USA. [2]Future Material Innovation Center, School of Materials Science and Engineering, Zhangjiang Institute for Advanced Study, and School of Physics and Astronomy, Shanghai Jiao Tong University, Shanghai 200240, China. [3]Department of Materials Design and Innovation, University at Buffalo, The State University of New York, Buffalo, NY, 14260, USA. [4]National Center for Electron Microscopy, Molecular Foundry, Lawrence Berkeley National Laboratory, Berkeley, CA 94720, USA. *Email: j.miao@ucla.edu*

**High-entropy alloys (HEAs) balance mixing entropy and intermetallic phase formation enthalpy, creating a vast compositional space for structural and functional materials[1-6]. They exhibit exceptional strength-ductility trade-offs in metallurgy[4-10] and near-continuum adsorbate binding energies in catalysis[11-16]. A deep understanding of crystal nucleation and growth in HEAs is essential for controlling their formation and optimizing their structural and functional properties. However, atomic-scale nucleation in HEAs challenges traditional theories based on one or two principal elements[17-23]. The intricate interplay of structural and chemical orders among multiple principal elements further obscures our understanding of nucleation pathways[5,24-27]. Due to the lack of direct three-dimensional (3D) atomic-scale observations, previous studies have relied on**

**simulations and indirect measurements[28-32], leaving HEA nucleation and growth fundamentally elusive. Here, we advance atomic electron tomography[33,34] to resolve the 3D atomic structure and chemical composition of 7,662 HEA and 498 medium-entropy alloy nuclei at different nucleation stages. We observe local structural order that decreases from core to boundary, correlating with local chemical order. As nuclei grow, structural order improves. At later stages, most nuclei coalesce without misorientation, while some form coherent twin boundaries. To explain these experimental observations, we propose the gradient nucleation pathways model, in which the nucleation energy barrier progressively increases through multiple evolving intermediate states. We expect these findings to not only provide fundamental insights into crystal nucleation and growth in HEAs, but also offer a general framework for understanding nucleation mechanisms in other materials.**

HEA nanoparticles composed of Co, Ni, Ru, Rh, Pd, Ag, Ir, and Pt were synthesized using a shock synthesis method[35]. Metal precursors with the desired multi-element compositions were rapidly heated to approximately 2000 K for 50 ms, followed by rapid cooling at around $10^5$ K/s, resulting in the emerging nuclei being kinetically trapped at room temperature (Methods). Tomographic tilt series were acquired from 25 HEA nanoparticles, labelled as HEA1-25, using a scanning transmission electron microscope in annular dark-field mode (Extended Data Table 1). After image pre-processing, each tilt series was reconstructed with an advanced iterative algorithm[36], and the 3D atomic coordinates were traced and refined, producing an experimental atomic model for each nanoparticle (Methods). The precision of the 3D atomic coordinates was determined to be 20 pm[37]. Figure 1a-c show three representative experimental atomic models for HEA2, HEA6, and HEA10, corresponding to early, mid, and late-stage nucleation, respectively.

Due to AET's limited ability to distinguish elements with small atomic number differences[37], the eight elements in the HEAs were classified into three types: Co and Ni as type 1 (blue); Ru, Rh, Pd, and Ag as type 2 (yellow); and Ir and Pt as type 3 (red) (Methods).

To quantify the local structural order, we calculated a normalized bond-orientational order (BOO) for each atom, where BOO = 0 indicates a completely disordered structure and BOO = 1 represents a perfect face-centred cubic (fcc) structure (Methods). Using a threshold of BOO $\geq$ 0.5 [27,37], we characterized the degree of crystallinity of the 25 HEA nanoparticles, which range from 22.2% to 99.9% (Extended Data Fig. 1). Figure 1d-f show the 3D distribution of BOO in HEA2, HEA6, and HEA10, with degrees of crystallinity of 26.8%, 55.2%, and 86.8%, respectively. At early-stage nucleation, most regions are disordered (Fig. 1d), with some crystalline features starting to form (blue) and very few regions exhibiting pronounced crystallinity (red). As nucleation proceeds, crystalline features become more abundant, and more regions develop pronounced crystallinity (Fig. 1e). At late-stage nucleation, most regions exhibit a well-defined crystalline structure (Fig. 1f). From the experimental atomic models, we computed the pair-distribution functions of the 25 HEA nanoparticles. These functions gradually transition from broad and amorphous features in HEA1 to sharp and crystalline features in HEA25 (Fig. 1g, bottom to top). Additionally, the short-range order was quantified using Voronoi tessellation[37,38]. By indexing the local polyhedron surrounding each atom with <$n_3$, $n_4$, $n_5$, $n_6$>, where $n_i$ is the number of $i$-fold symmetries, this method provides both the local symmetry and the coordination number around each atom. Figure 1h shows the 12 most abundant Voronoi polyhedra in HEA2 (green), HEA6 (blue), and HEA10 (red). The rapid increase of <0, 3, 6, 3> and <0, 4, 4, 4> from HEA2 to HEA10 indicates the formation of more crystalline structures.

**Capturing the 3D atomic structure of HEA nuclei at different nucleation stages**

To identify all the nuclei in the HEA nanoparticles, we used the following criteria: nuclei consist of atoms with BOO $\geq$ 0.5, each nucleus has the highest local BOO, and neighbouring nuclei are separated by boundaries with the lowest local BOO (Methods). We determined a total of 7,662 nuclei from the 25 HEA nanoparticles at different stages of nucleation. The smallest nuclei contain single core atoms with BOO $\geq$ 0.5, surrounded by atoms with BOO < 0.5 (Extended Data Fig. 2). Figure 2a-e shows five representative nuclei and their BOO distribution, containing 10, 96, 238, 613, and 1,290 atoms, respectively. These nuclei exhibit non-spherical shapes, with the highest BOO at the cores and the lowest BOO at the boundaries, while the cores are not necessarily located at the geometrical centre of the nuclei (Extended Data Fig. 3). As the nuclei grow, the regions with high crystallinity expand progressively (Fig. 2a-e). By statistically analysing all the nuclei, we observed that the number of nuclei exponentially decreases with increasing size (Fig. 2f). Additionally, the BOO of the nuclei cores is positively correlated with nuclei size (Fig. 2g).

Next, we quantitatively characterized the local crystallinity inside the nuclei by radially averaging the BOO of each nucleus. Figure 2h depicts the distribution of BOO relative to the distance from the nuclei core across five groups categorized by nuclei size: <10 (blue dots), 10-100 (red dots), 100-500 (yellow dots), 500-1000 (purple dots), and >1000 (green dots). Each nuclei size category corresponds to the number of atoms contained within each nucleus. A consistent trend observed is the gradual decrease in BOO from nuclei cores to boundaries. As nuclei increase in size, their core BOO becomes more crystalline, and the distribution of BOO extends further along the radial direction. To quantify these observations, we fit the BOO distribution with a generalized Gaussian distribution function, $\alpha(r) = \alpha_0 \times e^{-\left(\frac{r}{R}\right)^\beta}$, where $\alpha_0$, $R$ and $\beta$ are three fitting parameters. The solid curves in Fig. 2h show the fitting results, with the fitting parameters listed in Extended Data Table 2. These experimental results differ from both classical nucleation theory (CNT) and the two-step nucleation model. CNT assumes a uniform crystalline structure within each nucleus, with the nucleus radius as the sole variable[18–20], whereas the two-step model posits that a disordered cluster of solute molecules forms first, followed by the development of a crystal nucleus within the cluster[39–41].

**Gradient nucleation pathways**

To better explain our experimental results, we derived the gradient nucleation pathways (GNP) model (Methods),

$$\Delta G = -\int \Delta g\, \alpha(\vec{r})\, dV + \int \gamma \left| \vec{\nabla}\, \alpha(\vec{r}) \right| dV, \qquad (1)$$

where $\Delta G$ is the total free energy change, $\Delta g$ is the free energy change per unit volume, $\alpha(\vec{r})$ is the order parameter distribution between 0 and 1 (in this case, the BOO), and $\gamma$ is the interfacial tension of a sharp interface that can be anisotropic. Equation (1) applies to homogeneous nucleation, while for heterogeneous nucleation, it is multiplied by a shape factor[18,20]. The first term in Eq. (1) represents the effective volume energy change of a nucleus, while the second term accounts for its effective interfacial energy. The interfacial energy can be derived by dividing the nucleus into many small volumes and summing the contributions from all these small volumes (Extended Data Fig. 4). GNP generalizes CNT by allowing a gradual change in the order parameter at the interface, effectively lowering nucleation energy barriers (Extended Data Fig. 5). In the sharp interface limit, GNP reduces to CNT, demonstrating CNT as a special case within the broader GNP model. Using the experimental BOO distribution $\alpha(r)$ in Fig. 2H, we calculated $\Delta G$ as a function of the nucleus radius using Eq. (1), as depicted in Fig. 2i. We observed multiple evolving intermediate states, with the energy barrier gradually increasing as the nuclei grow. Additionally, we experimentally resolved the 3D atomic structures of 498 nuclei from six medium-entropy alloy (MEA) nanoparticles composed of Ni, Pd, and Pt (Extended Data Fig. 6a), and performed a quantitative analysis of their energetics. Extended Data Fig. 6b shows the radial average BOO distribution as a function of distance from the MEA nuclei core, with solid curves representing fits to a generalized Gaussian function. By applying the experimental data from Extended Data Fig. 6b to Eq. (1), we computed the total free energy change as a function of radial distance. The analysis revealed a gradual increase in the nucleation energy barrier (Extended Data Fig. 6c), consistent with the HEA results (Fig. 2i).

To corroborate our experimental findings, we conducted ab initio molecular dynamics (AIMD) simulations of HEAs containing the same eight atomic species (Methods). These simulations captured atomic dynamics during solidification, providing insights into the structural evolution essential for crystal nucleation. An analysis of 4,591 simulated nuclei revealed an exponential decrease in the number of nuclei as their size increased (Extended Data Fig. 7a). The radial average BOO distribution from the nuclei core is shown in Extended Data Fig. 7b, with solid curves fitted to a generalized Gaussian function. Applying this data to Eq. (1), we calculated the total free energy change as a function of radial distance, confirming that the nucleation energy barrier gradually increases across multiple evolving intermediate states (Extended Data Fig. 7c). We also tracked the evolution of a nucleus as the temperature decreased from 1,300 K to 700 K.

Extended Data Fig. 7d and e show the radial average BOO distribution during the growth of the nucleus and the corresponding total free energy change, respectively, revealing a gradual increase in the nucleation energy barrier. These simulations are consistent with our experimental observations and further support GNP. Compared to CNT[18–20] and the two-step nucleation model[39–41], GNP not only provides a better explanation for our results, but also lows the energy barrier through multiple intermediate states.

**Correlation between nucleation and local chemical order**

To study the impact of local chemical order on nucleation in HEAs, we quantified the chemical short-range order (CSRO) parameters, $\alpha_{ij}$, for each atom based on the elemental species of its nearest neighbors[25] (Methods). A positive $\alpha_{ii}$ value indicates the segregation of atomic species $i$, whereas a negative value implies the opposite. For $i \neq j$, a negative $\alpha_{ij}$ value denotes intermixing between atomic species $i$ and $j$, while a positive value signifies the opposite. Figure 3a and b show 3D volume renderings of $\alpha_{33}$ and BOO in HEA10, exhibiting local heterogeneity in both CSRO and BOO. To better visualize the correlation between nucleation and CSRO, Figure 3c–f present two perpendicular slices of $\alpha_{33}$ and BOO. These slices demonstrate that regions with significant type 3 segregation, i.e., clusters of type 3 elements, coincide with areas of enhanced crystallinity. Statistical analysis of all the 7,662 HEA nuclei confirms a positive correlation between $\alpha_{33}$ and BOO (Fig. 3g), revealing that nucleation in HEAs tends to initiate in type 3 segregated regions. Further analysis of the other CSRO parameters, presented in Extended Data Fig. 8, reveals that nucleation is unfavourable in regions with type 12 intermixing, i.e., intermixing between type 1 and type 2 elements. To examine the generality of the correlation between CSRO and nucleation, we performed the same analysis on 498 nuclei from six NiPdPt MEA nanoparticles. Our findings confirm that crystal nucleation preferentially occurs in Pt-segregated regions, while it is suppressed in NiPd intermixing regions (Extended Data Fig. 9).

These experimental observations can be attributed to the following physical mechanisms. Type 3 elements (i.e., Ir, Pt) have high atomic numbers, resulting in stronger atomic interactions, reduced atomic mobility, and localized stabilization of the crystal structure. These properties make type 3 segregated regions energetically favourable sites for nucleation. In contrast, type 12 intermixing regions, which involve elements with lower atomic numbers (e.g., Co, Ni, Ru, Rh, Pd, Ag), are associated with greater chemical disorder and variability in atomic interactions. This increased disorder raises the system's entropy, making such regions less favourable for the formation of a crystalline nucleus. These findings highlight the important role of local chemical order in nucleation within HEA and MEA nanoparticles, a phenomenon likely to be general across multi-principal element materials.

**Nuclei coalescence and twin formation**

As nucleation progresses, smaller nuclei grow and coalesce. To elucidate this growth mechanism, we calculated the orientation of each nucleus using a template matching method (Methods) and quantified the misorientation angle ($\theta$) between pairs of nuclei. Figure 4a shows the distribution of nuclei pairs as a function of $\theta$ and the pair separation, revealing three distinct groups with $\theta$ around 0º, 37º, and 60º. When the pair separation approaches zero, the majority of nuclei pairs cluster around 0º with some centring around 60º, while the number of nuclei pairs around 37º is reduced. To further illustrate this point, we used stereographic triangle plots to depict the rotation axes for the nuclei pairs with separations of 2 nm and 0 nm, as shown in Fig. 4b and c, respectively. For the aligned nuclei pairs ($\theta \approx 0$º), the pair orientations are uniformly distributed in the stereographic plots and are independent of the pair separation. When the pair separation approaches zero (i.e., nuclei coalescence), nuclei pairs with $\theta \approx 37°$ cluster around the [110] rotation axis (Fig. 4c, middle). Those with $\theta \approx 60°$ primarily aggregate around the [111] rotation axis, with a small fraction around the [110] axis (Fig. 4c, right).

To quantitatively characterize the growth mechanism, we employed the coincidence site lattice method[42,43] to determine the boundary orientations of the coalescing nuclei with respect to crystallinity. We calculated the sigma ($\Sigma$) values for the boundaries of the coalescing nuclei pairs, where the $\Sigma$ value measures the degree of coincidence between two crystal lattices along a boundary (Methods). Figure 4d presents histograms of the boundary orientations of the coalescing nuclei in 25 HEA nanoparticles, which are categorized into three degrees of crystallinity: <50%, 50-95%, and >95%. The predominant boundaries are $\Sigma1$, indicating that the majority of nuclei coalesce with no misorientation. Figure 4e shows two coalescing nuclei with a $\Sigma1$ boundary. The second most abundant boundaries are $\Sigma3$, characterized by a 60° rotation about the [111] axis, resulting in coherent twin formation. Figure 4f illustrates a representative $\Sigma3$ boundary between two coalescing nuclei.

In addition to $\Sigma1$ and $\Sigma3$, we also observed a minimal number of $\Sigma9$, $\Sigma11$, $\Sigma19$, and $\Sigma27$ boundaries, as well as a small fraction of randomly oriented boundaries (Fig. 4d, inset). Among these high $\Sigma$ value boundaries, $\Sigma9$ has the largest fraction, resulting from the intersection of two $\Sigma3$ boundaries. Figure 4g illustrates the coalescence of three nuclei characterized by two $\Sigma3$ boundaries and one $\Sigma9$ boundary. As the degree of crystallinity of the HEA nanoparticles changes from <50% to >95%, the fraction of $\Sigma1$ boundaries increases from 66.1% to 90.8%, while the $\Sigma3$ boundaries decrease from 17.8% to 8.1%, and the other high $\Sigma$ value boundaries reduce to minimal (Fig. 4d). These experimental results illustrate the mechanisms of nucleation growth and twin formation at the atomic scale, complementing in situ transmission electron microscopy studies of nanocrystal growth and coalescence[43].

**Conclusions**

We employed AET to determine the 3D atomic structure and chemical composition of 7,662 HEA nuclei and 498 MEA nuclei at different stages of nucleation. We observed that the nuclei exhibit the highest structural order at their cores, with crystallinity gradually decreasing towards the boundaries. As the nuclei grow, their population decreases exponentially, while their cores become more crystalline. By quantifying the BOO gradient from the cores to the boundaries of the nuclei, we found that the distribution of the BOO gradient extends further along the radial direction as the nuclei size increases. Since these observations are inconsistent with CNT[18-20] or the two-step nucleation model[39-41], we formulated the GNP model to better explain our experimental results. By incorporating our experimental data into the equation, we observed that GNP exhibits multiple evolving intermediate states and a gradual increase in the nucleation energy barrier as nuclei grow. This demonstrates that GNP captures the energetics of nucleation more accurately than CNT and the two-step model. Furthermore, by analysing BOO and CSRO in HEA and MEA nuclei, we identified a correlation between nucleation and local chemical order. At the late stage of nucleation, the majority of nuclei coalesce without misorientation, while a small fraction forms coherent twin boundaries. These findings provide a fundamental understanding of crystal nucleation and growth in HEAs, which could guide the rational design of HEAs for applications in catalysis and metallurgy. Finally, although the GNP model was derived from analysing a large number of HEA and MEA nuclei, we expect it has broad applicability for understanding a wide range of nucleation processes[44].

**Figures and Figure legends**

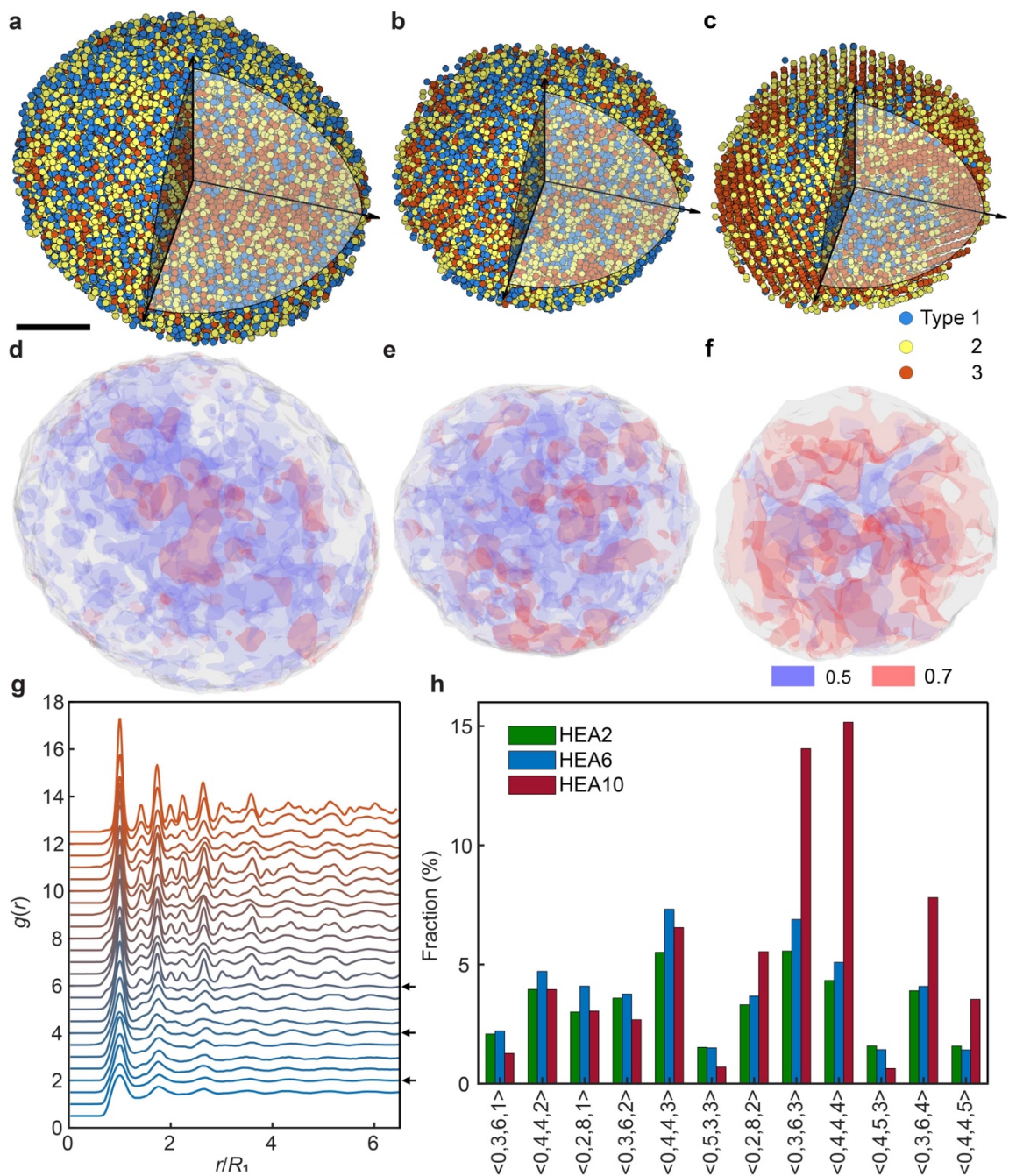

**Fig. 1 | 3D atomic structure and chemical composition of 25 HEA nanoparticles at different stages of nucleation. a-c**, Experimental 3D atomic models of HEA2, HEA6, and HEA10, illustrating early, mid, and late-stage nucleation, respectively. **d-f**, 3D iso-surface renderings of the local structural order in HEA2, HEA6, and HEA10, showing

BOO values of 0.5 (blue) and 0.7 (red), respectively. **g**, Pair-distribution functions of the 25 HEAs, with a gradual increase in structural order from HEA1 (bottom) to HEA25 (top). The three arrows indicate HEA2, HEA6, and HEA10. **h**, Twelve most abundant Voronoi polyhedra in HEA2 (green), HEA6 (blue), and HEA10 (red), shown with increasing coordination numbers. Scale bar, 2 nm.

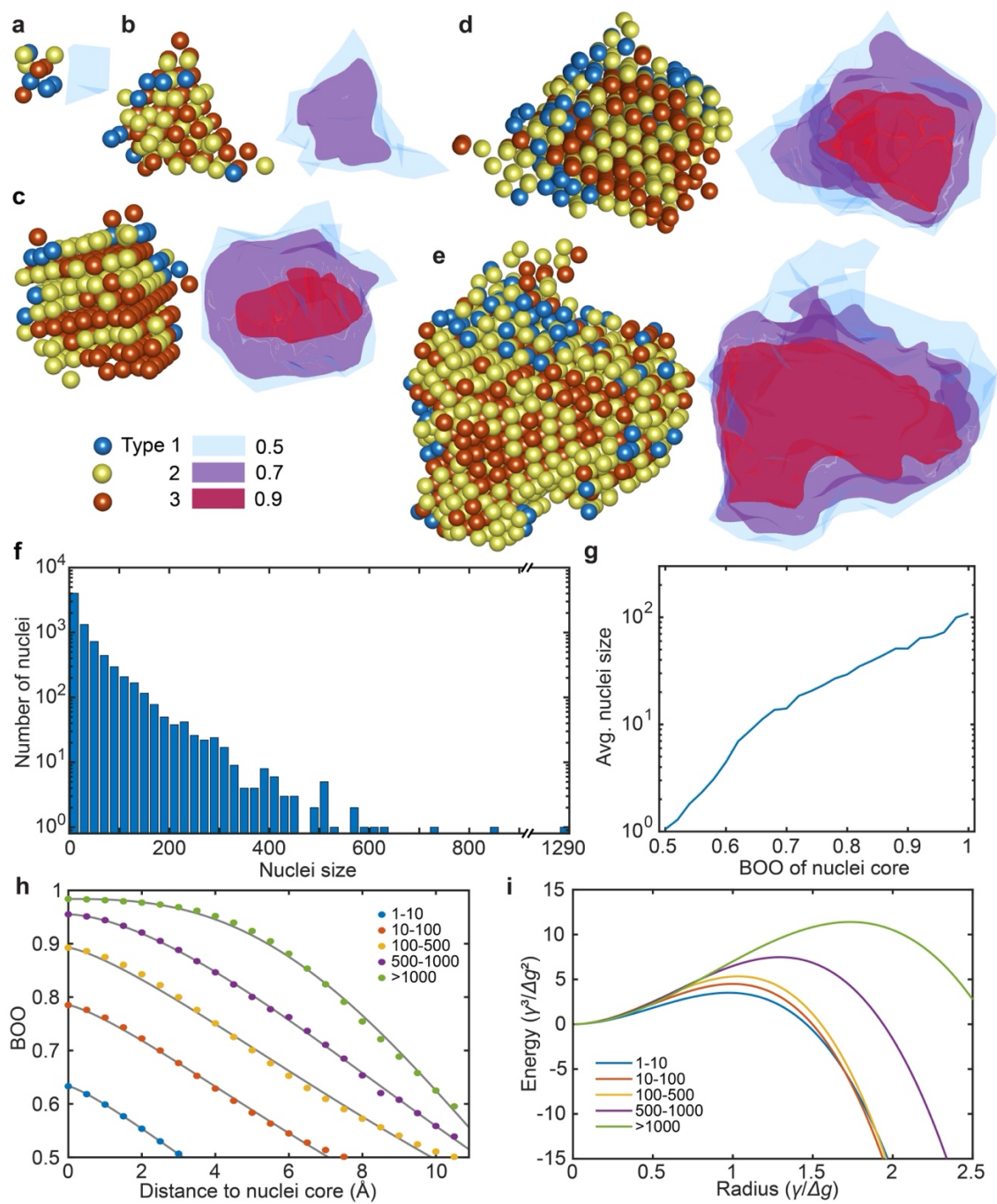

**Fig. 2 | Gradient nucleation pathways. a-e,** Five representative nuclei and their 3D BOO distributions, containing 10, 96, 238, 613, and 1,290 atoms, respectively. Iso-surface renderings display BOO values of 0.5 (blue), 0.7 (purple), and 0.9 (red). **f,** Exponential decrease in the number of nuclei as a function of their size. **g,** Positive correlation between the BOO of the nuclei cores and the size of the nuclei. **h,** Radial average BOO distribution as a function of the distance to the nuclei core in five groups of varying nuclei sizes: <10 (blue dots), 10-100 (red dots), 100-500 (yellow dots), 500-1,000 (purple dots), and >1,000 (green dots). The solid curves represent fits to a generalized Gaussian distribution function. **i,** Total free energy change as a function of radial distance, calculated by applying the experimental data from (**g**) into Eq. (1).

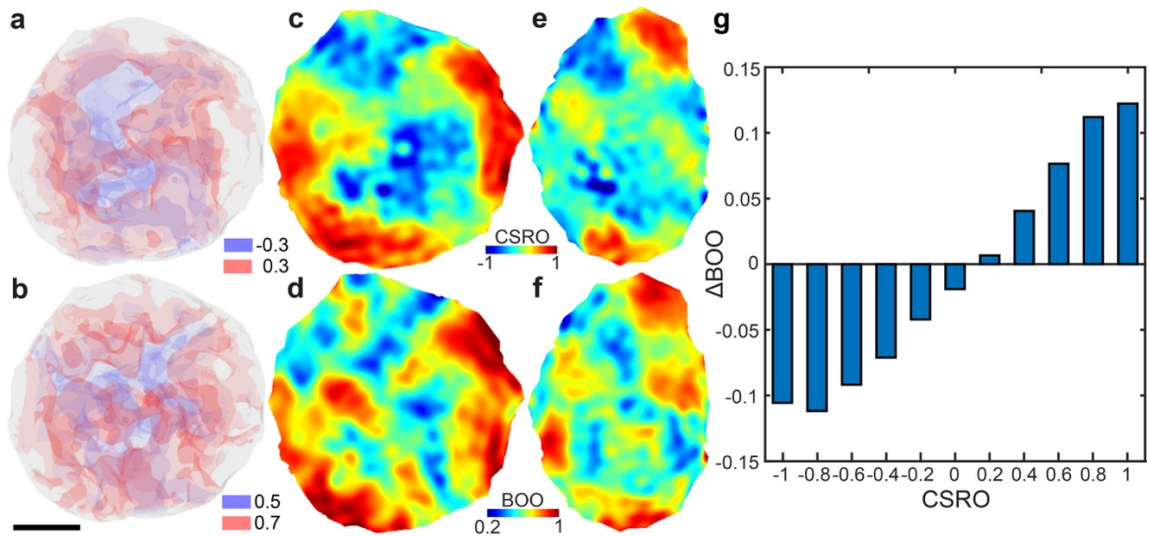

**Fig. 3 | Correlation between nucleation and local chemical order. a, b,** 3D volume renderings of $\alpha_{33}$ (**a**) and BOO (**b**) in HEA10, illustrating the correlation between the local chemical and structural order. **c-f,** Two perpendicular slices in the XY (**c, d**) and XZ (**e, f**) planes of CSRO ($\alpha_{33}$) and BOO in HEA10, respectively, showing regions with strong type 3 segregation correlate with areas of high structural order. **g,** The increase in BOO as a function of $\alpha_{33}$, obtained by averaging the 25 HEA nanoparticles, further

confirming that high structural order favours type 3 segregation. The other CSRO parameters are illustrated in Extended Data Fig. 8. Scale bar, 2 nm.

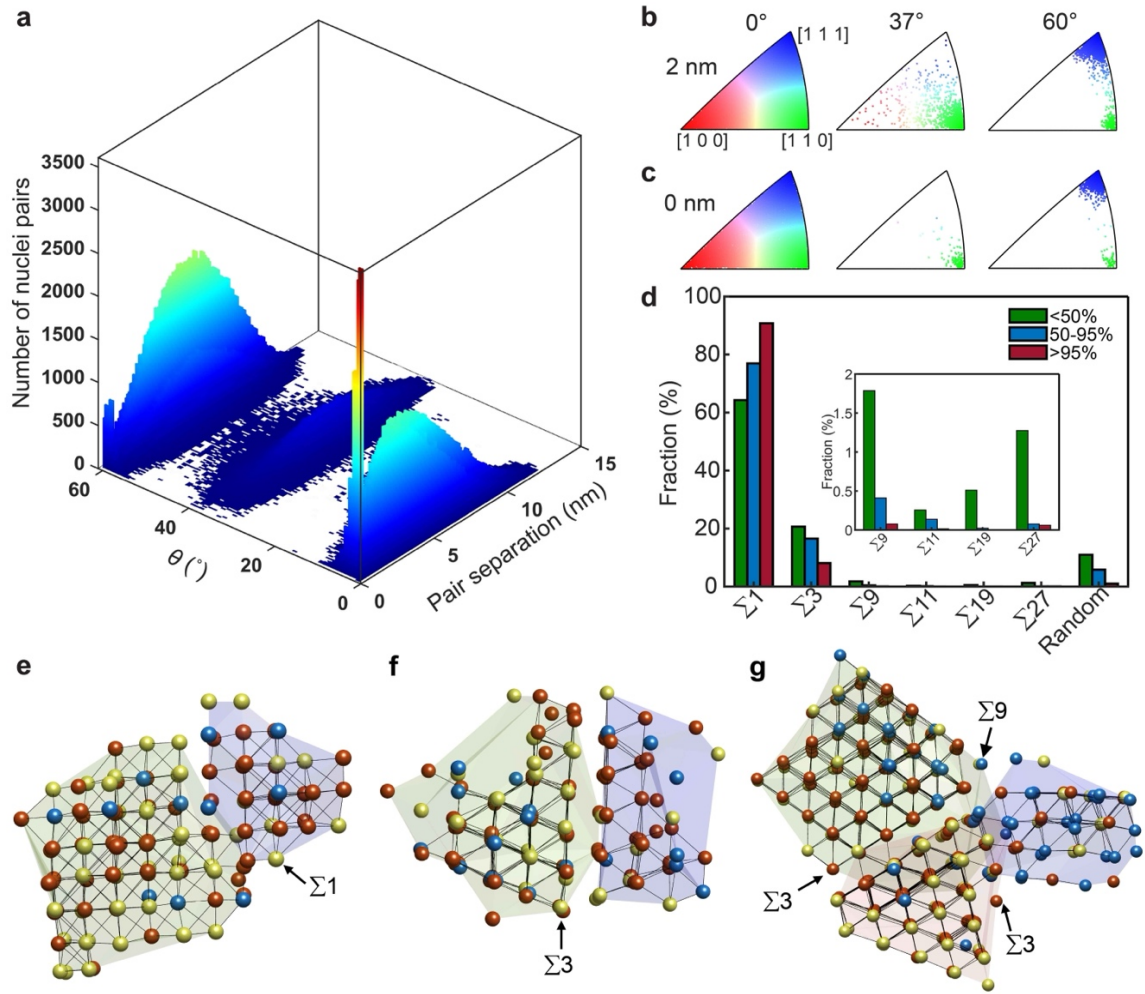

**Fig. 4 | Nuclei coalescence and twin formation. a**, Distribution of nuclei pairs as a function of the misorientation angle ($\theta$) and the pair separation, showing three distinct groups with $\theta$ around 0º, 37º, and 60º. **b**, **c**, Stereographic triangle plots of the rotation axes of nuclei pairs with separations of 2 nm and 0 nm, showing $\theta$ values around 0º, 37º, and 60º. The red, green, and blue dots represent nuclei pairs near the [100], [110], and [111] rotation axes, respectively. **d**, Histograms of the $\Sigma$ boundaries of the coalescing nuclei in the 25 HEA nanoparticles. The nanoparticles are categorized into three different degrees of crystallinity: <50% (green), 50-95% (blue), and >95% (red). **e**, Two coalescing

nuclei with no misorientation (i.e., forming a Σ1 boundary). **f**, A representative Σ3 boundary between two coalescing nuclei, resulting in twin formation. **g**, The coalescence of three nuclei, featuring two Σ3 boundaries and one Σ9 boundary.

## METHODS

### HEA nanoparticle synthesis and preparation

The HEA nanoparticles were synthesized using the carbothermal shock method[35]. Initially, metal precursors (Co, Ni, Ru, Rh, Pd, Ag, Ir, Pt) were dissolved in ethanol solvent (0.05 M) to achieve the desired composition. These mixed precursors were then dispersed and stabilized on a reduced graphene oxide substrate. After air-drying at ambient temperature, the precursors on the substrate underwent rapid heating to about 2,000 K for 50 ms using Joule heating, followed by rapid cooling at approximately $10^5$ K/s. This fast cooling process allowed us to kinetically trap HEA nanoparticles at various stages of nucleation.

### AET data acquisition

The experimental AET tilt series of 25 HEA nanoparticles was acquired using the TEAM 0.5/TEAM 1 microscope and TEAM stage at the National Center for Electron Microscopy. The aberration-corrected microscope operated in annular dark-field (ADF) mode at an accelerating voltage of 200/300 kV (Extended Data Table 1). For each nanoparticle, sample features in its vicinity were identified at each tilt angle to aid in nanoparticle localization and optimize imaging conditions, thereby minimizing unnecessary exposure to the electron beam. To correct for drift distortion, three to four consecutive images were acquired using the same fast scanning direction and a dwell time of 3 μs. To minimize electron damage, the beam current was optimized, resulting in a total electron dose of $3.4$-$10.2 \times 10^5$ e⁻/Å² for each tilt series (Extended Data Table 1). Images captured before, during, and after the acquisition of each tilt series confirmed that the structural changes of each nanoparticle remained minimal throughout the data collection process.

### Image pre-processing

The AET tilt series underwent a detailed multi-step preprocessing protocol prior to tomographic reconstruction, outlined as follows.

i) Drift correction. Multiple images taken consecutively at each tilt angle were used to determine the drift vector through cross-correlation algorithms. A shift step size of 0.1 pixel was employed during the calculation to detect subpixel drift in the images. Using the drift vectors, a 2D mesh grid with corrected pixel positions for each image was computed. The raw images were then interpolated onto the drift-corrected 2D mesh grids and averaged to produce a single drift-corrected image at each angle.

ii) Image denoising. The block-matching and 3D filtering (BM3D) algorithm was employed to remove Poisson and Gaussian noise from the drift-corrected images[45]. To optimize the BM3D algorithm parameters, we first estimated the Poisson and Gaussian noise levels in the experimental images. Subsequently, the same noise levels were added to simulated ADF-scanning transmission electron microscopy (STEM) images of a nanoparticle with a similar size and composition to the experimental samples. The optimum parameters were identified by achieving the best cross-correlation between the noise-free images and their denoised versions. Once optimized, these parameters were applied to the experimental images for effective noise reduction.

iii) Background subtraction. To perform background subtraction on the denoised images, a mask slightly larger than the nanoparticle boundary was generated for each image using the Otsu thresholding method. The background beneath the nanoparticle was then estimated using Laplacian interpolation from the boundary of the mask. This estimated background was subsequently subtracted from the denoised image[37].

iv) Image alignment. To align the images in each tilt series, sharing a common tilt axis, the centre of mass positions were computed for each image. Each individual image was then shifted so that its centre of mass coincided with the centre of the image. This procedure ensured that all images in the tilt series were properly aligned and ready for AET reconstruction[46,47].

**AET reconstructions**

After pre-processing, each tilt series was reconstructed using the Real Space Iterative REconstruction (RESIRE)[36] algorithm with an oversampling ratio of 4. Convergence was achieved over 200 iterations to ensure accuracy. To compensate for nominal tilt angle errors and minor misalignments, an iterative process for angular refinement and spatial alignment was employed[37]. Following the initial 3D reconstruction, a finer mask of the HEA nanoparticle was projected at each tilt angle, which was used to update the background subtraction and the pre-processed tilt series. The final 3D reconstruction was then obtained using the RESIRE algorithm with further angular refinement and spatial alignment.

**3D atomic structure and chemical composition of the 25 HEA nanoparticles**

The 3D atomic structure and chemical composition of the HEA nanoparticles were obtained by tracing atom positions and classifying atom species using the final reconstructions.

i) To trace the 3D coordinates of the atoms, each reconstruction was oversampled by a factor of 3 using spline interpolation. Local maxima were then identified and fit using a polynomial function within a vicinity volume of 0.8×0.8×0.8 Å$^3$, generating a preliminary list of potential atoms[37,48]. Considering the typical atomic radii in HEAs, a minimum distance of 2 Å between neighbouring atoms was enforced to help rule out unphysical atom locations. Non-atoms, typically originating from weak noise in the reconstruction, were removed by K-means clustering of the integrated intensities around potential atoms[49,50]. The remaining atoms were manually checked against the reconstruction slice by slice to correct for misidentified or unidentified atoms, which typically constituted a small fraction (<1%) of the total atom population in each HEA nanoparticle.

ii) Using the traced atom positions, species classification within each HEA nanoparticle was conducted in two stages. First, K-means clustering was applied to the integrated intensity over a 0.8×0.8×0.8 Å$^3$ volume around each atom, resulting in an initial classification into three types (Co, Ni as type 1; Ru, Rh, Pd, Ag as type 2; Ir, Pt as type 3). This initial classification was followed by a local classification, where the averaged intensities of type 1-3 atoms within a 10 Å radius around a centre atom were computed[23,37]. The type of each centre atom was then determined by the minimum $L_2$-norm between its intensity and the averaged intensities of the three types. After iterating through all atoms, the final classification of the three types was obtained.

iii) By fitting and assigning each type of atom with a Gaussian function, the 3D coordinates of the atoms were further refined by minimizing the $L_2$-norm error between calculated and experimental images using a gradient descent algorithm[50]. The final atomic structures of the HEA nanoparticles with refined 3D coordinates and species were used for further analysis. To ensure the accuracy of the further analysis on the atomic structures, surface atoms of each HEA nanoparticle were excluded prior to computing structural and chemical orders.

**Bond orientational order (BOO) and crystallinity analysis**

The atomic scale structural order during nucleation was quantified using normalized BOO, based on the averaged local bond orientational order parameters, $Q_4$ and $Q_6$[51]. The normalized BOO is defined as $\frac{\sqrt{Q_4^2+Q_6^2}}{\sqrt{Q_{4fcc}^2+Q_{6fcc}^2}}$, where $Q_{4fcc}$ and $Q_{6fcc}$ are the reference values for the fcc lattice. A BOO value of 1 indicates a perfect local fcc arrangement around the corresponding atom, while 0 indicates a random arrangement. A BOO cutoff of 0.5 was chosen as the threshold for separating crystalline and amorphous atoms. The crystallinity of each HEA nanoparticle was defined as the fraction of its constituent crystalline atoms. In our analysis, we did not detect hexagonal close-packed (hcp) nuclei within the HEA nanoparticles. However, hcp-ordered atoms were found localized along $\Sigma 3$ twin boundaries formed between adjacent fcc nuclei.

**Identification and orientation analysis of nuclei**

The nuclei within each HEA nanoparticle were identified using the following procedure. For every atom in the nanoparticle, a sphere was drawn with a radius equivalent to the first nearest neighbour distance, determined by the position of the first valley in the atom's pair distribution function. If all the atoms within this sphere had a BOO equal to or lower than that of the considered atom, these atoms were grouped in the same cluster. Conversely, if any atom in the sphere had a higher BOO than the considered atom, the considered atom was assigned to the cluster of the atom with the highest BOO within the sphere. After iterating through all atoms, clusters where the BOO of constituent atoms was equal to or greater than 0.5 were identified as nuclei. It is important to note that no size constraints were imposed on the nuclei, allowing for the identification of early-stage metastable nuclei, which could be as small as a single atom.

The orientation of each nucleus was determined by fitting the 3D coordinates of its constituent atoms to a fcc reference lattice using a brute-force template matching approach. A cutoff distance of 15% of the fcc lattice parameter was used to decide whether an atom belongs to the template, based on its deviation from the reference position in the template. To avoid arbitrary orientation choices for very small nuclei consisting of only a few atoms, only nuclei with a minimum of 13 atoms in their fcc template were considered to have well-defined orientations and were used for further analysis.

**Chemical short-range order (CSRO) analysis**

In our AET experiments, ADF-STEM images were acquired for each nanoparticle, followed by the reconstruction of the 3D atomic structure. The integrated intensity of each atom in the 3D reconstruction is proportional to its atomic number. For the MEA nanoparticles, we successfully identified the atomic species as Ni, Pd, and Pt. However, for the HEA nanoparticles, the eight constituent elements were classified into three groups due to minimal differences in their atomic numbers: Co and Ni as type 1, Ru, Rh, Pd, and Ag as type 2, and Ir and Pt as type 3. The CSRO ($\alpha_{ij}$) of each atom was computed by analyzing the chemical species of its nearest neighbor atoms using the following equation[25,52],

$$\alpha_{ij} = \frac{p_{ij} - C_j}{\delta_{ij} - Cj} \qquad (2)$$

where $p_{ij}$ denotes the probability of finding a $j$-type atom among the nearest neighbours of an $i$-type atom, $C_j$ represents the concentration of $j$-type atoms in the nanoparticle, and $\delta_{ij}$ is the Kronecker delta function.

**Determining the boundary orientations of merging nuclei pairs**

The orientation of each nucleus can be described by a matrix consisting of three lattice vectors of its fcc template. The relative rotation matrix of a nuclei pair was calculated by comparing their orientation matrices, which was then converted to a rotation angle, $\theta$, about a rotation axis, $\boldsymbol{a}$. Due to the equivalence of the six <100> directions of the fcc template, permuting these lattice vectors results in multiple choices of ($\theta$, $\boldsymbol{a}$) for each nuclei pair. To eliminate ambiguity, the permutation with the minimum rotation angle, $\theta$, was selected, representing the minimum misorientation between the nuclei pair.

According to the coincidence site lattice theory[42], the boundaries of nuclei pairs can be categorized using the notation $\Sigma n$, where $n$ denotes the ratio of the unit cell volume of the coincidence site lattice to the unit cell volume of the original crystal lattices. Reference values of ($\theta_n$, $\boldsymbol{a}_n$) for each $\Sigma n$ were obtained from the literature[42]. The experimental nuclei boundaries were classified by comparing their misorientation ($\theta$, $\boldsymbol{a}$) with the reference values for each $\Sigma n$, using a maximum angle difference of 5° between the rotation axes $\boldsymbol{a}$ and $\boldsymbol{a}_n$, and an angular error of $|\theta - \theta_n| \leq \theta_m$, where $\theta_m = 15°/\sqrt{n}$ as defined by the Brandon criterion[53].

**Ab initio molecular dynamics simulations**

AIMD simulations provide insights into atomic dynamics during solidification, offering detailed information on the chemical interactions of the involved species. In this study, AIMD simulations based on density functional theory were performed using the Vienna *ab initio* simulation package (VASP)[54-56]. The exchange and correlation interactions were described using the generalized gradient approximation (GGA) functional of Perdew-Burke-Ernzerhof (PBE)[57], and the Brillouin zone was sampled at the gamma point. The simulation cell, measuring 5×5×4 fcc conventional cells and containing 400 atoms randomly assigned to represent an equimolar HEA, was employed with periodic boundary conditions. Prior to AIMD simulations, the structures were relaxed using a conjugate gradient algorithm with a force convergence criterion of 0.01 eV/Å[58]. AIMD simulations were then conducted in the NVT ensemble with a time step of 2 fs[59,60]. To generate a liquid/amorphous HEA configuration, the system was heated to and thermalized at 2,300 K for 4 ps. Subsequently, the system was cooled to room temperature (300 K) at a cooling rate of 1

K/ps. A temperature segmentation strategy was implemented, dividing the cooling process into separate simulations performed in parallel over 100 K intervals[61,62]. This segmented AIMD method effectively captures the structural evolution during solidification by addressing the limited long-range diffusion inherent to constrained simulation time scales while significantly optimizing computational resources.

To examine the nucleation process observed between 1,300 K and 700 K (identified in the segmented simulations) in greater detail, a continuous cooling simulation spanning this temperature range was performed. The parameters of the continuous simulation were kept consistent with those used in the previous parallel simulations, with the canonical ensemble and Nose-Hoover thermostat employed. It is important to note that the cooling rate in this single continuous solidification simulation was relatively higher to enhance computational efficiency.

**Derivation of the GNP model**

To derive GNP as described by Eq. (1) in the main text, we assign each atom or molecule in a nucleus an order parameter ($\alpha$) that ranges from 0 to 1. Summing the order parameters for all atoms, the first term in Eq. (1) represents the effective volume energy difference of the nucleus. For example, an atom with $\alpha = 0.6$ contributes $-0.6 \Delta g \Delta V$ to the effective volume energy difference, where $\Delta V$ is the volume occupied by the atom.

To derive the second term in Eq. (1), we divide the nucleus into numerous small volumes, each with an area $\Delta s$ and a width $\Delta d$ (Extended Data Fig. 4). The direction of the order parameter gradient within each volume aligns with the $\Delta s$ direction, while the magnitude of the gradient is $|\Delta \alpha|/\Delta d$, where $\Delta \alpha$ is the order parameter difference within the volume. The interfacial tension for this volume is calculated as:

$$\frac{|\Delta \alpha| / \Delta d}{1 / \Delta d} \gamma = |\Delta \alpha| \gamma \qquad (3)$$

where $\gamma$ is the interfacial tension of a sharp interface with $\Delta \alpha = 1$. The total interfacial energy of the nucleus is obtained by summing the interfacial energies of all the small volumes:

$$\int |\Delta \alpha| \gamma \, ds = \int \gamma \left| \frac{\Delta \alpha}{\Delta d} \right| \, dV = \int \gamma |\vec{\nabla} \alpha| \, dV \qquad (4)$$

which corresponds to the second term in Eq. (1).

GNP differs fundamentally from the squared-gradient model, which incorporates a squared-gradient term, $\left( \vec{\nabla} c \right)^2$, where $c$ may represent density, composition, or order parameters[20,63]. The distinctions between the squared-gradient term and the magnitude-gradient term , $|\vec{\nabla}\alpha|$, in GNP can be summarized as follows.

1) *Derivation*. The squared-gradient term arises from a Taylor expansion, as the gradient term vanishes when summed across all directions[20,63]. In contrast, the magnitude-gradient term is derived through a multi-step process: (i) dividing a nucleus into small volumes (Extended Data Fig. 4); (ii) normalizing the interfacial tension of each volume relative to that of a sharp interface; and (iii) integrating the interfacial energy over all volumes. Consequently, the magnitude-gradient term is of a lower order than the squared-gradient term.

2) *Coefficients*. While both terms can utilize the order parameter, their coefficients differ. The squared-gradient term requires a complex integration to determine its coefficient[20,63]. On the other hand, the magnitude-gradient term's coefficient, $\gamma$, has a clear physical interpretation as the interfacial tension of a sharp interface where $\alpha$ transitions between 1 and 0.

3) *Relationship to CNT*. Substituting the Heaviside step function (as shown in Eq. (5)) into the magnitude-gradient term in Eq. (1) reduces GNP to CNT, making CNT a special case of GNP. In contrast, the squared-gradient model cannot be reduced to CNT even when the Heaviside step function is applied.

**Nucleation energy barriers in GNP**

To further elucidate GNP, we calculate the total free energy change as a function of radial distance for a sharp interface and three diffuse interfaces using Eq. (1). It is important to emphasize that the order parameter distribution, $\alpha(\vec{r})$, in Eq. (1) does not inherently assume a radial distribution within a nucleus, as $\vec{r}$ represents a 3D vector. However, to provide clear physical insights into GNP, we adopt a radial distribution for $\alpha(r)$ in our analysis.

1) *Sharp interface*. A sharp interface is represented by the Heaviside step function (Extended Data Fig. 5a, black curve):

$$\alpha = \begin{cases} \alpha_0 & r \leq R \\ 0 & r > R \end{cases} \qquad (5)$$

where $\alpha_0 \leq 1$ is a constant. Substituting Eq. (5) into Eq. (1) gives:

$$\Delta G = -\frac{4\pi}{3}\alpha_0 R^3 \Delta g + 4\pi\alpha_0 R^2 \gamma \qquad (6)$$

With $\alpha_0 = 1$, Eq. (6) corresponds to CNT. By differentiating $\Delta G$ with respect to $R$, we calculate the critical radius ($r_c^*$) and the nucleation energy barrier ($\Delta G^*$) for the sharp interface:

$$r_c^* = \frac{2\gamma}{\Delta g} \qquad \Delta G^* = \frac{16.8\gamma^3}{\Delta g^2} \qquad (7)$$

2) *Diffuse interface: Linear decrease.* A diffuse interface with a linear decrease in $\alpha$ (Extended Data Fig. 5a, red curve) is represented by:

$$\alpha = \alpha_0 \left(1 - \frac{r}{R}\right) \qquad (8)$$

Substituting Eq. (8) into Eq. (1) and differentiating $\Delta G$ with respect to $R$, we derive:

$$r_c^* = \frac{2.7\gamma}{\Delta g} \qquad \Delta G^* = \frac{9.9\alpha_0\gamma^3}{\Delta g^2} \qquad (9)$$

3) *Diffuse interface: Parabolic decrease (above linear).* A parabolic decrease in $\alpha$ (above the linear line in Extended Data Fig. 5a, blue curve) is specified as:

$$\alpha = \alpha_0 - \alpha_0 \left(\frac{r}{R}\right)^2 \qquad (10)$$

Substituting Eq. (10) into Eq. (1) yields:

$$r_c^* = \frac{2.5\gamma}{\Delta g} \qquad \Delta G^* = \frac{13.1\alpha_0\gamma^3}{\Delta g^2} \qquad (11)$$

4) *Diffuse interface: Parabolic decrease (below linear).* A parabolic decrease in $\alpha$ (below the linear line in Extended Data Fig. 5a, green curve) is represented by:

$$\alpha = \alpha_0 \left(\frac{r}{R} - 1\right)^2 \qquad (12)$$

Substituting Eq. (12) into Eq. (1), we obtain:

$$r_c^* = \frac{3.3\gamma}{\Delta g} \qquad \Delta G^* = \frac{7.8\alpha_0\gamma^3}{\Delta g^2} \qquad (13)$$

Extended Data Fig. 5b illustrates the total free energy change as a function of radial distance, calculated using the Heaviside step function (black), the linear function (red), and the two parabolic functions (blue and green) in Extended Data Fig. 5a via Eq. (1). As the order parameter distribution extends further radially, the nucleation energy barrier gradually increases from the green to the red, blue, and black curves in Extended Data Fig. 5b.

**Data availability**

All the raw and processed experimental data will be immediately posted on an open-access repository (https://github.com) after the paper is published online.

**Code availability**

All the MATLAB source codes for the 3D reconstruction, atom tracing, and data analysis of this work will be immediately posted on an open-access repository (https://github.com) after the paper is published online.

**Acknowledgements** This work was primarily supported by STROBE: A National Science Foundation Science and Technology Center under grant number DMR 1548924. The ADF-STEM imaging with TEAM 0.5 was performed at the Molecular Foundry, which is supported by the Office of Science, Office of Basic Energy Sciences of the US DOE under contract no. DE-AC02-05CH11231.